\documentstyle[psfig,12pt]{article}
\begin{document}
\baselineskip = 14pt

\title{Subharmonic Motion of Particles \\
in a Vibrating Tube}
\author{Jysoo Lee \\
\\
Department of Physics, Seoul National University \\
Seoul 151-742, Korea}

\date{Feb 23, 1998}
\maketitle

\begin{abstract}

We study the motion of strongly inelastic particles in a narrow
vibrating tube using molecular dynamics simulation.  At low frequency
of the vibration, we observe qualitative changes of the motion, as the
depth of the pile increases.  The center of mass of the particle cloud
can be described by a superposition of modes of different frequencies.
For certain values of the depth, a single mode dominates.  The
frequency of the dominant mode is $1/2$, $1/3$, or $1/1$ of the
vibration.  We suggest that the behavior can be understood in terms of
a new time-scale $\tau$, reflecting the recompaction time for a
finite-depth pile.

\vspace*{20pt}


\end{abstract}

\newpage


Granular material under vibration has been a constant source of
interesting phenomena, such as convection, heap formation and surface
waves \cite{c90,hb92,mb94,hnst95,jnb96}.  When the depth is
sufficiently small (a shallow bed), the motion of a pile under
vibration is qualitatively the same as that of a single layer.  As the
depth increases, changes are expected, which raises several
interesting questions.  For example, at what depth does the motion
start to significantly differ from that of a single layer?  Do the
changes occur gradually or suddenly with increase of the depth?

These questions are also closely related to some of proposed
mechanisms for parametric waves and convection cells observed in
vibrated granular material.  For example, an estimate of the onset of
the parametric waves, given in Ref.~\cite{mus94}, is in excellent
agreement with experiment.  The estimation is based on the motion of a
single particle.  Also related is an argument for convection, which
involves a bifurcation from a single particle motion \cite{hyh95}.  A
systematic study of the depth dependence of the motion is necessary to
check the validity and limitations of such mechanisms.


There have been several studies on the state of granular material
under vibration.  Cl\'{e}ment and Rajchenbach measured the density,
velocity and temperature fields of two-dimensional packing of beads
\cite{cr91}.  Luding {\it et al.} studied the density field of the one
and two-dimensional systems, and found a new scaling \cite{lhb94},
which was confirmed by Warr {\it et al.}  \cite{whj95} and Lee
\cite{l95}.  Lan and Rosato measured the density and temperature
fields of the three-dimensional system \cite{lr94}.

While these works focus on the dependence of the fields on the
frequency and amplitude of the vibration, there are few works
specifically on the effect of the depth of a pile.  Thomas {\em et
al.}  studied the system in three dimensions, focusing on a shallow
bed \cite{tmls89}, and found four distinct behaviors as the depth
increases.  They are (1) ``Newtonian-I,'' where the particles are
bouncing so randomly that there is little change in the density field
during a cycle, (2) ``Newtonian-II,'' where a dense layer of particles
forms during one part of each cycle, (3) ``coherent-expanded,'' where
the particles move as a coherent mass, and (4) ``coherent-condensed,''
where the particles move as a coherent mass but remain compact through
a cycle.  Also, Brennen {\it et al.} studied the effect of the depth
as well as the amplitude on the expansion of a pile \cite{bgw96}.
They found at least one sudden change in the expansion at certain
amplitude, whose value changes with the depth.


In this paper, we systematically study the depth dependence for
various combinations of the amplitude and frequency of the vibration.
We use the molecular dynamics (MD) simulation method, which provides
detailed information on the motion of individual particles as well as
time averaged fields.

In the experiment of Thomas {\it et al}, the motion of a pile
approaches that of a single block, as the depth increases
\cite{tmls89}.  We are interested to know whether that is the only
possible dependence.  Here, we choose to use strongly inelastic
particles.  By doing so, we can study the motion in the other extreme,
since typical value of $e$ used in experiments is rather large
\cite{kwg97}.  Here, $e$ is the coefficient of restitution between the
particles. Furthermore, one may argue that a pile with large $e$
behaves like a shorter pile with small $e$, since a pile with large
$e$ can be divided into blocks, each of which acts like a single
particle.

We find that the system shows rich and unexpected dependence on the
depth.  In general, several modes of different frequencies are
necessary to describe the resulting motion.  However, at specific
values of the depth, one of them dominates.  The dominant mode is,
besides single particle motion, always a subharmonic ($1/2$ or $1/3)$
of the frequency of vibration.  We suggest that these behaviors result
from an additional recompaction time-scale $\tau$ introduced in the
system.  When an initially compact pile is launched from the bottom,
it takes time $\tau$ to be compact again after subsequent collisions.
If $\tau$ becomes comparable to or even larger than the period $T$ of
the vibration, we expect the motion of the pile to be significantly
different from that of a single layer.  A dominant mode seems to
occur, when $\tau$ is close to an integer multiple of $T$.  At
somewhat higher frequency ($100$ Hz), a pile moves as a single block,
and $\tau$ becomes negligible.



The simulations are done in two dimensions with disk shaped particles,
using a form of interaction due to Cundall and Strack
\cite{cs79,lh93}.  Particles interact only by contact, and the force
between two such particles $i$ and $j$ is the following.  Let the
coordinate of the center of particle $i$ ($j$) be $\vec{R}_i$
($\vec{R}_j$), and $\vec{r} = \vec{R}_i - \vec{R}_j$.  The normal
component $F_{j \to i}^{n}$ of the force acting on particle $i$ from
particle $j$ is
\begin{equation}
\label{eq:fn}
F_{j \to i}^{n} = k_n (a_i + a_j - \vert \vec{r} \vert) 
                - \gamma m_e (\vec{v} \cdot \hat{n}),
\end{equation}
where $a_i$ ($a_j$) is the radius of particle $i$ ($j$), and $\vec{v}
= d\vec{r}/dt$.  Here, $k_n$ is the elastic constant, $\gamma$ the
friction coefficient, and $m_e$ is the effective mass, $m_i m_j/(m_i +
m_j)$.  The shear component $F_{j \to i}^{s}$ is given by
\begin{equation}
F_{j \to i}^{s} = - {\rm sign} (\delta s) ~ 
{\rm min}(k_s \vert \delta s \vert, \mu \vert F_{j \to i}^n \vert),
\label{eq:fs}
\end{equation}
where $\mu$ is the friction coefficient, $\delta s$ the {\em total}
shear displacement during a contact, and $k_s$ the elastic constant of
a virtual tangential spring.  The shear force applies a torque to the
particles, which then rotate.

Particles can also interact with walls.  The force and torque on
particle $i$, in contact with a wall, are given by (\ref{eq:fn}) -
(\ref{eq:fs}) with $a_j = 0$ and $m_e = m_i$.  Also, the system is in
a vertical gravitational field $\vec{g}$.  The interaction parameters
used in this study are fixed as follows, unless otherwise specified:
$k_n = k_s = 5 \times 10^4, \gamma = 10^3$ and $\mu = 0.2$.  In order
to avoid artifacts of a monodisperse system (e.g., hexagonal packing),
we choose the radius of the particles from a Gaussian distribution
with the mean $0.1$ and width $0.02$.  The density of the particles is
$5$.  Throughout this paper, CGS units are implied.


We put the particles in a two-dimensional rectangular box.  The box
consists of two horizontal (top and bottom) plates which oscillate
sinusoidally along the vertical direction with given amplitude $A$ and
frequency $f$.  The width and height of the box is $1$ and $10^4$,
respectively.  The small width is used to suppress the surfaces waves
\cite{mus94}.  We apply a periodic boundary condition in the
horizontal direction.


The coefficient of restitution between the particles $e_{pp}$,
determined from the above interaction parameters, is $8.0 \times
10^{-2}$, and the coefficient between the particles and the wall is
$e_{pw}$ $2.5 \times 10^{-3}$.  The particles are thus almost
completely inelastic.  We have studied the motion of a single particle
for several values of $A$ with $f = 10$, and find good agreements with
the predictions of Mehta and Luck \cite{ml90}.


We measure the time series $Y(t)$ of the center of mass of the
particles, and compare it with that of a single particle.  In Fig.~1,
we show the power spectrum $P(f)$ of the series for several values of
$H$.  Here, $f = 10, \Gamma \equiv A (2 \pi f)^{2} / g = 2$, and $H$,
the total number of layers in a pile, is $1, 8, 12, 16, 28$ (from
bottom to top).  The measurements are made for $200$ cycles.  The
motion of a single particle with these parameters is known to have the
same period as the vibration \cite{gh83}.  This is confirmed by the
fact that $P(f)$ with $H = 1$ is strongly dominated by the mode at $f
= 10$.

For larger values of $H$, however, the behavior becomes quite
different.  As $H$ increases, the $f = 10$ mode becomes less dominant
($H \sim 8$), then an $1/2$ subharmonic mode becomes dominant ($H \sim
12$).  By further increasing $H$, no clear dominant mode is present
($H = 16$), and an $1/2$ subharmonic mode dominates again ($22 < H <
30$).  Thus, several modes are always present in the spectrum, and one
of them dominates around specific values of $H$.  We find that these
qualitative features of the power spectrum seem to be insensitive to
small changes of the width, coefficients of restitution and elastic
constant.  



We now investigate the mechanism for the behaviors.  We start with the
observation that the particles in a box do not remain as a single
block, but tend to be dispersed.  We introduce quantity $R(t)$, which
characterizes the dispersion, as $R(t) \equiv \sqrt{\langle y^2
\rangle - \langle y \rangle ^2}$.  Here $y_i(t)$ is the vertical
coordinate of particle $i$, and the average is taken over the
particles.  We describe the motion of the pile using their effective
center $Y(t)$ and effective ``radius'' $R(t)$.


The radius $R(t)$ does vary with time.  It remains small, while the
particles are resting on the bottom.  When they are launched into air,
$R(t)$ initially increases, then decreases after they collide with the
bottom.  This procedure introduces additional time-scale $\tau$---time
needed for an compact pile to be compact again after launching and
subsequent landing(s).  One can think of $\tau$ as ``relaxation time''
for the pile.  When $\tau$ becomes comparable to or even larger than
$T$, we expect that the motion of the pile can no longer be described
by that of a single particle.

To make the idea more quantitative, we measure $\tau$.  From the time
series $R(t)$, we locate the times at which $R(t)$ reaches local
minima.  The interval between successive minima is defined to be
$\tau$.  One should consider only the flights of initially compact
piles.  Some of the local minima correspond to partially expanded
state, and should not be used.  To take that into account, I discard
values of $\tau$ when $R(t)$ at the launching is larger than $20\%$ of
the maximum $R(t)$.  We calculate distribution $D(\tau)$ from the
resulting set of $\tau$.  The resulting $D(\tau)$ seems to be
insensitive to small variation of the cutoff.


When the motion of the pile is periodic, the particles are launched at
specific phase of the vibration.  One thus expects that $D(\tau)$
consists of a few sharp peaks.  When the motion becomes more chaotic,
the particles are launched at more irregular phase.  The relaxation
time $\tau$ becomes more random, and the peaks of $D(\tau)$ broaden
out.  Thus, one can think of $D(\tau)$ as a representation of $P(f)$
in temporal domain.  Sharply peaked $P(f)$ corresponds to sharply
peaked $D(\tau)$, broad $P(f)$ to broad $D(\tau)$.


We now quantify the dominance of a single mode.  We define
\begin{equation}
Q^2  \equiv {1 \over \tau _{max} - \tau _{min}} ~ \int_{\tau
_{min}}^{\tau _{max}} (D(\tau) - \bar{D})^2 d\tau,
\end{equation}
where
\begin{equation}
\bar{D} \equiv {1 \over \tau _{max} - \tau _{min}} ~
\int_ {\tau _{min}}^{\tau _{max}} D(\tau) d\tau.
\end{equation}
Here, $\tau_{min}$ ($\tau_{max}$) is the minimum (maximum) of measured
$\tau$.  The quantity $Q$ measures deviation from a uniform
distribution.  If the motion of the pile is periodic, $D(\tau)$
consists of a few sharp peaks, and $Q$ is large.  For a chaotic
motion, we expect small $Q$.  In the top part of Fig.~2, we show $Q$
measured for the parameters of Fig.~1.  The curve consists of three
peaks forming a ``w'' shape.  The number of the peaks, and their
locations ($H = 1, 12, 28)$, are what are expected from Fig.~1.  Note
that the system shows broad resonances around $H=28$.  Such quality of
agreement seems to be typical, which demonstrates the value of $Q$.
We also calculate the relative contributions of $P(f_o)$ and
$P(f_o/2)$ to the power specturm $P(f)$, where $f_o$ is the driving
frequency.  These contributions also peak around the same locations of
the peaks of $Q$.


The key quantity which can be calculated from $D(\tau)$ is
$\bar{\tau}$ \cite{note}.  At the bottom of Fig.~2, we show
$\bar{\tau}$ measured using the parameters of Fig.~1.  One can see
that $\bar{\tau}$ is indeed larger than the period of the vibration $T
= 0.1$ for most $H$.  Furthermore, notice that the dominances of a
single mode seem to occur when $\bar{\tau}$ is close to an integer
multiple of $T$, besides small $H$ (single particle motion).  The
dominance near $H = 28$ occurs when $\bar{\tau}$ is close to $2 T$,
where the frequency of the dominant mode is also $2 T$.  However, the
dominance near $H = 12$ occurs near, but not exactly at, where
$\bar{\tau}$ becomes $2 T$ ($H \sim 10$).


Based on these observations, we propose a possible mechanism for the
dominance.  When the pile, launched from the bottom, comes back to
collide with the bottom, there is still significant dispersion, and
the pile becomes compact again only after $\bar{\tau}$.  When
$\bar{\tau}$ is an integer multiple of $T$, we expect the pile to
repeat the same sequence of motions.  Thus, $\bar{\tau} = n T$ is a
condition for a single-mode dominance, and the period of the motion is
$n T$.  


The proposed mechanism gets further support by studying the motion for
different values of $\Gamma$.  In Fig.~3, we show $Q$ and $\bar{\tau}$
measured for $\Gamma = 3$ while the other parameters remain the same.
One can see four dominances: one particle motion at $H = 1$, two $1/2$
subharmonic ($H = 6$, and around $26$) motions, and an $1/3$
subharmonic ($H = 14$) motion.  Not only the observed periods ($1/1,
1/2, 1/3$), but also the order they appear ($1/1, 1/2, 1/3$ and $1/2$,
as $H$ increases) seems to be rather complex.  These can be easily
understood by looking at the corresponding $\bar{\tau}$.  The value of
$\bar{\tau}$ is indeed close to $2 T$ near $H = 6, 26$, where an $1/2$
subharmonic mode dominates.  However, there is only one dominance of
$1/3$ subharmonic mode near $H = 14$, compared to two expected from
$\bar{\tau}$ ($H \sim 12$ and $16$).  It is possible that the two
nearby dominances merge to form a single one.  As $\Gamma$ increases
to $4$ and $5$, all the dominances can again be explained from
$\bar{\tau}$.  No dominance with period larger than $3 T$ is observed.
As $\Gamma$ is further increased, the dominance becomes much less
clear.  


We also study the effect of $f$.  We change $f$ to $20$ and $100$
while keeping the other parameters fixed.  For $f = 20$ and $\Gamma =
2$, $Q$ assumes a ``w'' shape as in Fig.~2.  The dispersion of the
file for $f = 20$ is smaller than that of $f = 10$ at the same
$\Gamma$.  The decrease is more significant for $f = 100$, where the
systematic variation of $R(t)$ is too small to be seen.  This decrease
is not unexpected.  The expansion of the pile was shown to scale as $A
f$ rather than $\Gamma$ \cite{lhb94,whj95,l95}.  Therefore, the value
of $Af$ decreases, as $f$ increases for fixed $\Gamma$.


We argue that most aspects of the seemingly complex depth dependence
can be explained in terms of $\bar{\tau}$.  However, there are a few
remaining questions.  First, the origin of the discrepancy on the
location of dominances remains unclear.  Second, it is not clear why
only $1/n$ subharmonic mode with integer $n$ is observed.  In
principle, $\bar{\tau}$ can be made suitable for, e.g., a $2/3$ mode
by carefully tuning $H$.  


Finally, we discuss the conditions for observing the subharmonic
motion in an experiment.  The amplitude and frequency used in the
experiments of Thomas {\it et al.} are comparable to our simulations
\cite{tmls89}, but their results are quite different.  The difference,
we believe, is the coefficient of restitution.  The typical value of
$e$ used in the present simulations is less than $0.1$, while that of
a typical experiment is larger than $0.8$ \cite{kwg97}.  In order to
check the idea, we repeat the simulation with $e \sim 0.8$ for $f =
10$ and $\Gamma = 2$.  The motion of the pile is indeed chaotic at
small depth, and it becomes more coherent at larger depth, just as in
the experiments.  It is possible that the motion of a pile with small
$e$ is similar to that of a taller pile with larger $e$.  In such a
case, subharmonic motion could be observed in a taller pile with large
$e$.


We thank Joel Koplik and Yoon H. Hwang for useful discussions and
critical reading of the manuscript.  This work is supported in part by
the Department of Energy under grant DE-FG02-93-ER14327.  One of us
(J.L.) is supported in part by SNU-CTP and Korea Science and
Engineering Foundation through the Brain-Pool program.

\newpage
\begin{description}

\centerline{\psfig{figure=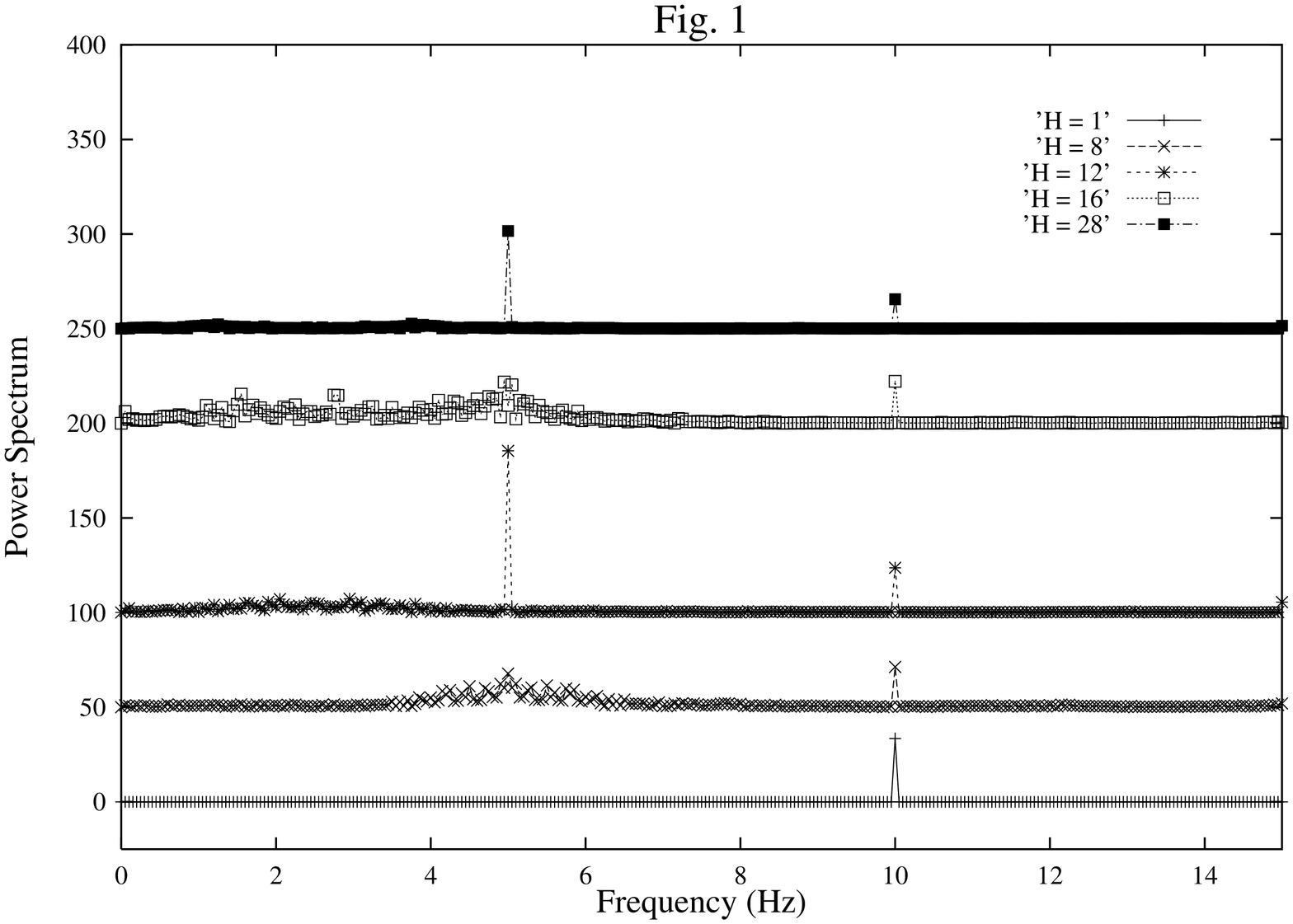,angle=270,width=5in}}

\item [Fig.~1:] Power spectra $P(f)$ of the center of mass of the
particles.  The five curves corresponds to, from bottom to top, $H =
1, 8, 12, 16, 28$.  The curves have been offset for clarity.  Here, $f
= 10$ and $\Gamma = 2$.
\vfill
\newpage

\centerline{\psfig{figure=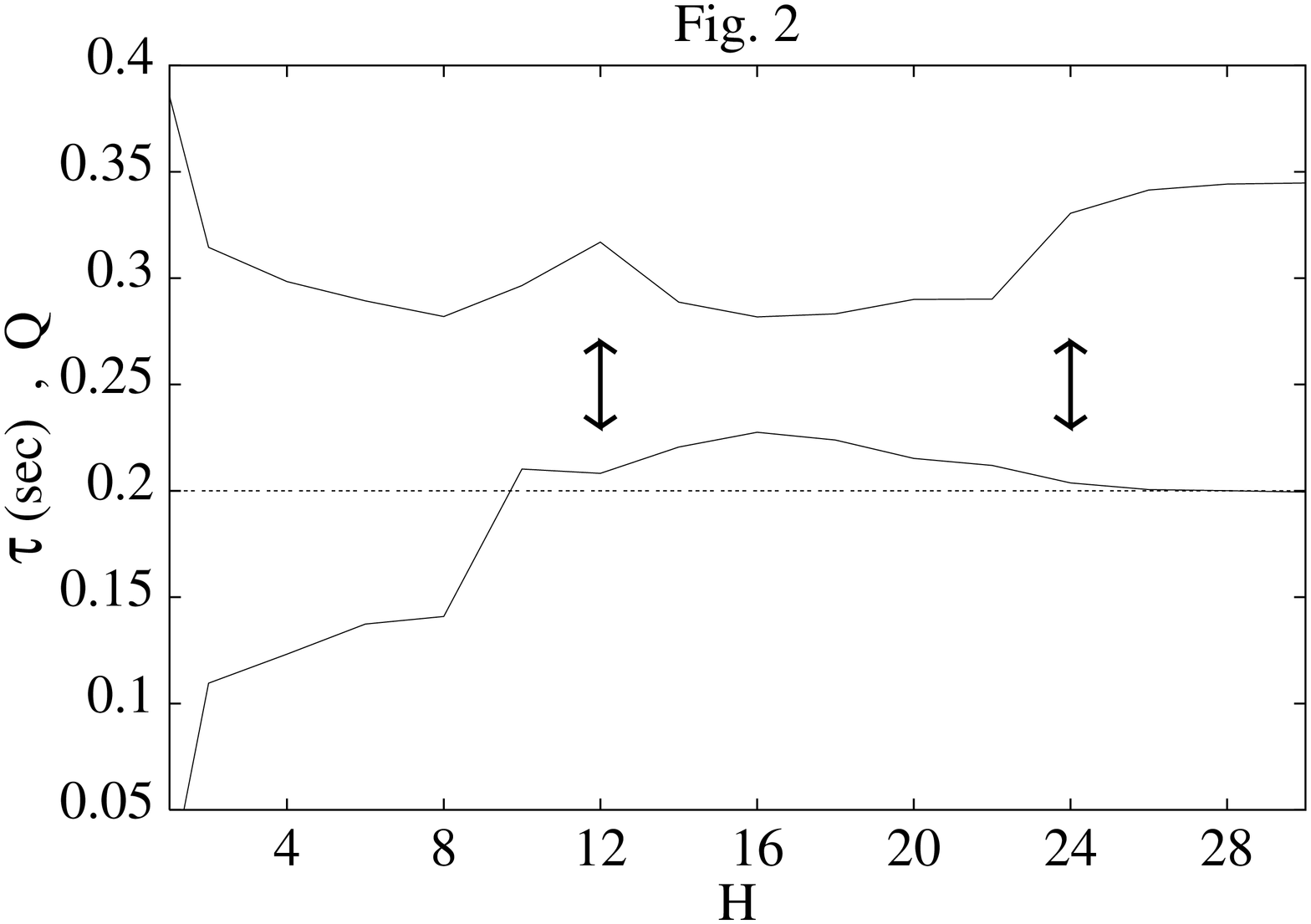,angle=270,width=5in}}

\item [Fig.~2:] The quantity $Q$ (top) which characterizes the
dominance of a single mode and the relaxation time $\bar{\tau}$
(bottom) are shown for several values of $H$.  The curve of $Q$ has
been offset and rescaled for clarity.  Parameters are identical to
those of Fig.~1.  There are three peaks in $Q$ at $H = 1$ (single
particle motion), $12$ ($1/2$ subharmonic), and broad peak near $H =
28$ ($1/2$ subharmonic).  Note that $\bar{\tau}$ is close to $2 T$
when subharmonic modes dominate.

\newpage

\centerline{\psfig{figure=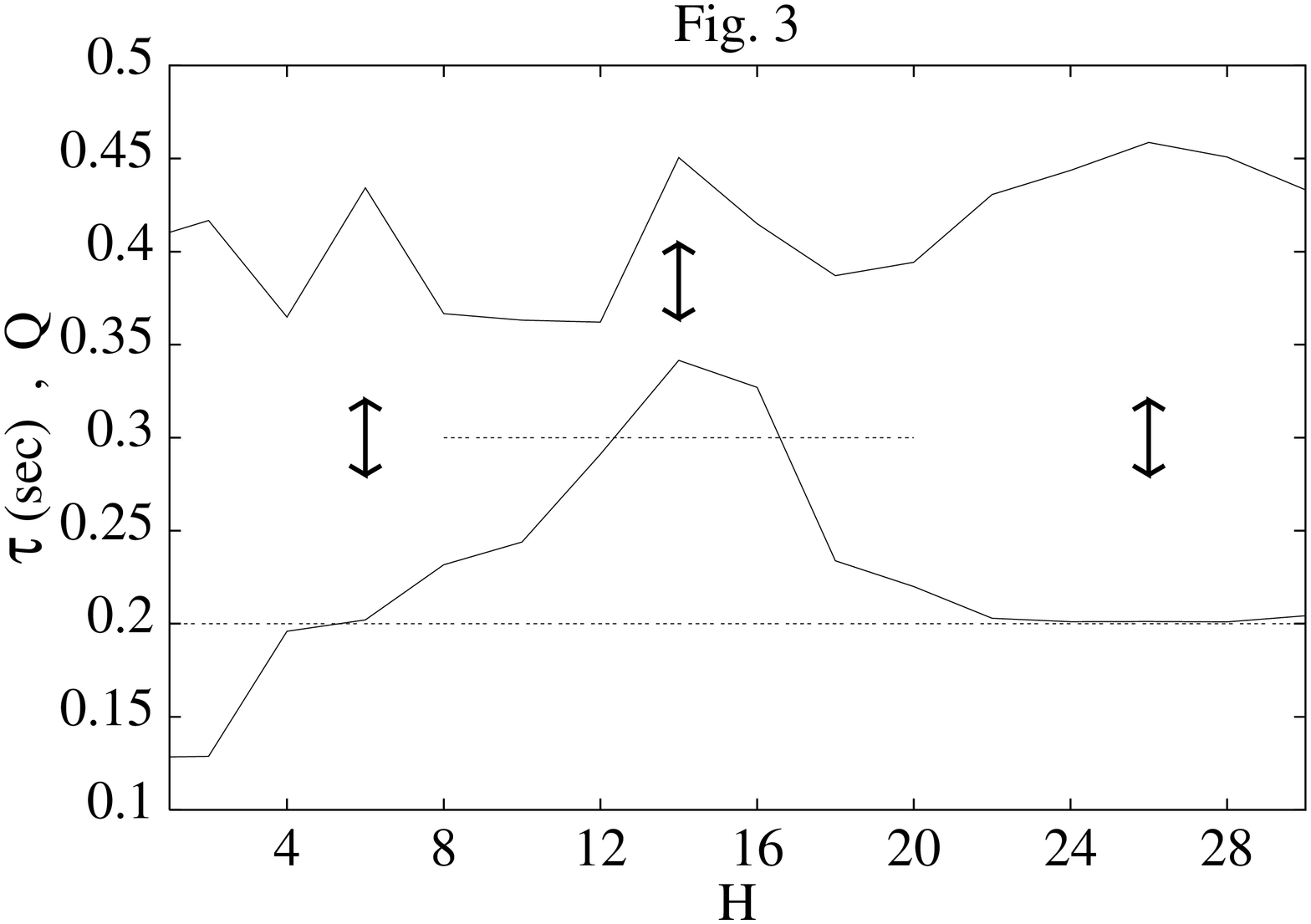,angle=270,width=5in}}

\item [Fig.~3:] The quantities $Q$ (top) and $\bar{\tau}$ (bottom) are
shown for $\Gamma = 3$ .  Other parameters are the same as Fig.~2.
There are four peaks in $Q$ at $H = 1$ (single particle motion), $6$
($1/2$ subharmonic), $14$ ($1/3$ subharmonic) and around $26$ ($1/2$
subharmonic).  Note that $\bar{\tau}$ is close to $2 T$ ($3 T$) when
an $1/2$ ($1/3$) subharmonic mode dominates.

\end{description}


\begin{thebibliography}{99}
\bibitem{note} We define $\bar{\tau} = \tau _o + 2 \delta \tau$, where
$\tau _o, \delta \tau$ is the average and standard deviation of the
dominant peak of $D(\tau)$, respectively.  We choose $\bar{\tau}$
larger than $\tau _o$ in order to give more weight for the
trajectories of initially compact piles.  Increasing the difference
from the mean ($2 \delta \tau$) results in small but systematic
increase of $\bar{\tau}$.
\end{thebibliography}
\end{document}